\documentclass[twocolumn,aps,prb,superscriptaddress,longbibliography]{revtex4-2}
\usepackage[colorlinks=true, citecolor=blue, urlcolor=blue, linkcolor=red]{hyperref}
\renewcommand{\section}[1]{{\par\it #1.~~}\ignorespaces}
\usepackage{amsmath,amssymb,tikz,scalerel}
\usepackage{cases}
\usepackage{bm}
\usetikzlibrary{svg.path}
\definecolor{orcidlogocol}{HTML}{A6CE39}
\tikzset{orcidlogo/.pic={
		\fill[orcidlogocol] svg{M256,128c0,70.7-57.3,128-128,128C57.3,256,0,198.7,0,128C0,57.3,57.3,0,128,0C198.7,0,256,57.3,256,128z};
		\fill[white] svg{M86.3,186.2H70.9V79.1h15.4v48.4V186.2z}
		svg{M108.9,79.1h41.6c39.6,0,57,28.3,57,53.6c0,27.5-21.5,53.6-56.8,53.6h-41.8V79.1z M124.3,172.4h24.5c34.9,0,42.9-26.5,42.9-39.7c0-21.5-13.7-39.7-43.7-39.7h-23.7V172.4z}
		svg{M88.7,56.8c0,5.5-4.5,10.1-10.1,10.1c-5.6,0-10.1-4.6-10.1-10.1c0-5.6,4.5-10.1,10.1-10.1C84.2,46.7,88.7,51.3,88.7,56.8z};}}
\newcommand\orcid[1]{\href{https://orcid.org/#1}{\mbox{\scalerel*{\begin{tikzpicture}[yscale=-1,transform shape]\pic{orcidlogo};\end{tikzpicture}}{|}}}}

\begin{document}
\title{Point-gap topology in amorphous non-Hermitian quantum systems}
\author{Xue-Min Yang\orcid{0000-0002-6937-8402}}
\affiliation{School of Electronic Science and Engineering, Chongqing University of Posts and Telecommunications, Chongqing 400065, China}
\affiliation{Chongqing Key Laboratory of Dedicated Quantum Computing and Quantum Artificial Intelligence, Chongqing 400065, China}
\affiliation{Key Laboratory of Quantum Artificial Intelligence and New Materials,Chongqing}
\affiliation{Institute for Advanced Sciences, Chongqing University of Posts and Telecommunications, Chongqing 400065, China}
\author{Mu Zhou}
\affiliation{School of Electronic Science and Engineering, Chongqing University of Posts and Telecommunications, Chongqing 400065, China}
\affiliation{Chongqing Key Laboratory of Dedicated Quantum Computing and Quantum Artificial Intelligence, Chongqing 400065, China}
\affiliation{Key Laboratory of Quantum Artificial Intelligence and New Materials,Chongqing}
\affiliation{Institute for Advanced Sciences, Chongqing University of Posts and Telecommunications, Chongqing 400065, China}
\author{Deng-Feng Li}
\affiliation{School of Electronic Science and Engineering, Chongqing University of Posts and Telecommunications, Chongqing 400065, China}
\affiliation{Chongqing Key Laboratory of Dedicated Quantum Computing and Quantum Artificial Intelligence, Chongqing 400065, China}
\affiliation{Key Laboratory of Quantum Artificial Intelligence and New Materials,Chongqing}
\affiliation{Institute for Advanced Sciences, Chongqing University of Posts and Telecommunications, Chongqing 400065, China}
\author{Jian Li \orcid{0000-0003-2443-0139} }
\affiliation{School of Electronic Science and Engineering, Chongqing University of Posts and Telecommunications, Chongqing 400065, China}
\affiliation{Chongqing Key Laboratory of Dedicated Quantum Computing and Quantum Artificial Intelligence, Chongqing 400065, China}
\affiliation{Key Laboratory of Quantum Artificial Intelligence and New Materials,Chongqing}
\affiliation{Institute for Advanced Sciences, Chongqing University of Posts and Telecommunications, Chongqing 400065, China}
\author{Jia-Ji Zhu \orcid{0000-0002-1328-356X} }
\affiliation{School of Electronic Science and Engineering, Chongqing University of Posts and Telecommunications, Chongqing 400065, China}
\affiliation{Chongqing Key Laboratory of Dedicated Quantum Computing and Quantum Artificial Intelligence, Chongqing 400065, China}
\affiliation{Key Laboratory of Quantum Artificial Intelligence and New Materials,Chongqing}
\affiliation{Institute for Advanced Sciences, Chongqing University of Posts and Telecommunications, Chongqing 400065, China}
\author{Li Li}
\affiliation{School of Electronic Science and Engineering, Chongqing University of Posts and Telecommunications, Chongqing 400065, China}
\affiliation{Chongqing Key Laboratory of Dedicated Quantum Computing and Quantum Artificial Intelligence, Chongqing 400065, China}
\affiliation{Key Laboratory of Quantum Artificial Intelligence and New Materials,Chongqing}
\affiliation{Institute for Advanced Sciences, Chongqing University of Posts and Telecommunications, Chongqing 400065, China}
\author{Qiu-Chi Chen}
\affiliation{School of Electronic Science and Engineering, Chongqing University of Posts and Telecommunications, Chongqing 400065, China}
\author{Xin-Qi Xu}
\affiliation{School of Electronic Science and Engineering, Chongqing University of Posts and Telecommunications, Chongqing 400065, China}
\author{Hao-Peng Zhang}
\affiliation{School of Electronic Science and Engineering, Chongqing University of Posts and Telecommunications, Chongqing 400065, China}
\affiliation{Chongqing Key Laboratory of Dedicated Quantum Computing and Quantum Artificial Intelligence, Chongqing 400065, China}
\affiliation{Key Laboratory of Quantum Artificial Intelligence and New Materials,Chongqing}
\affiliation{Institute for Advanced Sciences, Chongqing University of Posts and Telecommunications, Chongqing 400065, China}
\author{Hong Wu\orcid{0000-0003-3276-7823}}\email{Contact author: wuh@cqupt.edu.cn}
\affiliation{School of Electronic Science and Engineering, Chongqing University of Posts and Telecommunications, Chongqing 400065, China}
\affiliation{Chongqing Key Laboratory of Dedicated Quantum Computing and Quantum Artificial Intelligence, Chongqing 400065, China}
\affiliation{Key Laboratory of Quantum Artificial Intelligence and New Materials,Chongqing}
\affiliation{Institute for Advanced Sciences, Chongqing University of Posts and Telecommunications, Chongqing 400065, China}

\begin{abstract}
Recent studies have revealed that not only does the correspondence between spectral winding numbers and skin modes break down in non-Hermitian systems, but the energy spectrum itself is highly sensitive to generic perturbations, system size, and boundary conditions. In amorphous non-Hermitian systems, where the positions of lattice sites are uncertain, the spectral instability becomes even more severe, making it difficult to identify stable topological edge states from the eigenvalue spectrum alone. To overcome this challenge, we introduce a correspondence between stable zero-mode singular states and mid-gap states of the energy spectrum in the thermodynamic limit. Because the singular value spectrum is highly robust against small perturbations and variation in size, topological edge states can be reliably probed via singular values even in finite-sized systems.  Based on the singular-value decomposition of the Hamiltonian, we construct a topological invariant in real space to characterize the associated topologically protected edge states. Our approach provides a general strategy for exploring point-gap topology in real space and redefine the non-Hermitian skin effect from a new perspective.

\end{abstract}
\maketitle

\section{Introduction}
Non-Hermitian Hamiltonians provide a natural framework for describing open quantum systems, classical wave systems with gain and loss, and quasiparticles with finite lifetimes. Motivated by potential applications in sensors, lasers, unidirectional edge transport, and directional amplification \cite{PhysRevLett.125.180403,PhysRevResearch.4.013113,topologicallaser,PhysRevB.105.235426,PhysRevB.107.085302,rao2026unidirectionallikeedgetransportinduced,Wanjura_2020}, topological phases in non-Hermitian systems have attracted widespread attention in recent years \cite{PhysRevB.84.205128,PhysRevLett.116.133903,PhysRevLett.118.045701,PhysRevLett.121.086803,PhysRevLett.120.146402,PhysRevB.98.205417,PhysRevLett.123.066404,PhysRevX.9.041015,PhysRevLett.123.206404,PhysRevB.103.L041115,PhysRevLett.129.053903,PhysRevLett.130.157201,PhysRevLett.130.066601,PhysRevLett.130.266901,Okuma_2023,PhysRevB.109.205142,9b46-d2ry,q3db-9zkj,2zx3-rs57,ma2025dynamically,wu2025topological,PhysRevA.100.032102,ghorashi2025topologicalrealityswitchbulkboundary,xwb5-d933,x9kx-v9d2}. A key feature of these systems is point-gap topology, a concept with no Hermitian counterpart \cite{PhysRevLett.125.126402,PhysRevLett.124.086801,Denner_2021,RevModPhys.93.015005,Zhou_2023,PhysRevLett.132.136401,PhysRevB.109.035418,6dhc-34vj,xr6m-115j,ljvt-w6hw,ll76-j2l5}. According to the bulk-boundary correspondence, the point-gap topology is linked to the non-Hermitian skin effect (NHSE), where an extensive number of bulk eigenstates  are localized
at the edges \cite{PhysRevLett.123.016805,PhysRevLett.125.126402,PhysRevLett.124.086801,PhysRevB.99.201103,PhysRevB.103.205205,PhysRevB.104.L241402,PhysRevB.103.L140201,PhysRevB.103.125411,PhysRevLett.128.223903,PhysRevB.106.035425,PhysRevA.105.023311,PhysRevB.108.195126,PhysRevB.108.L060204,PhysRevResearch.6.013213,PhysRevA.110.022222,PhysRevB.111.155121,PhysRevB.111.115415,7m2n-9vmf,PhysRevB.111.174305,PhysRevB.111.125156,LongwenZhou2026075202,q3db-9zkj,li2026tunablenonhermitianskineffect,1zkn-gw9g,xbzb-jgk9,ljvt-w6hw,2f5t-gc98}. This connection has established a new paradigm for topological phases, which has been experimentally demonstrated across various platforms \cite{PhysRevLett.115.040402,topologicallaser,Xiao_2020,PhysRevResearch.2.023265,Zhang_2021,Zou_2021,PhysRevLett.129.070401,PhysRevLett.129.070401,Gu_2022,PhysRevLett.131.207201,PhysRevResearch.5.043034,PhysRevLett.132.063804,jiang2024observationnonhermitianboundaryinduced,PhysRevLett.132.063804,Zhao_2025,PhysRevLett.134.176601,Shen_2025,wang2026nonhermitiananomalousscalingengineering,l54w-77kc,s4fz-41lg}. 

 Recent studies have shown that in translationally invariant one-dimensional systems, the connection between point-gap winding and the skin effect holds only for the single-band case and generally breaks down in the presence of multiple bands \cite{sirker2026pseudospectralphenomenaoriginnonhermitian}. More fundamentally, the eigenvalue of a non-Hermitian Hamiltonian is highly sensitive to perturbations, boundary conditions, and system size \cite{PhysRevLett.134.056601,f37j-l6wp,qy19-f9wm,Trefethen2005,SciPostPhys.21.1.013,g1cw-tk7f}.  Consequently, identifying stable topological edge states directly from the energy spectrum poses a fundamental challenge.

A promising solution to this issue was recently proposed in Ref.~\cite{PhysRevLett.134.056601}, which utilized the singular value decomposition (SVD) of the Hamiltonian to probe topological zero modes.  They
have established a correspondence between stable zero-mode singular states and stable zero-mode topological
edge states in energy spectrum for the thermodynamic
limit. Since the singular spectrum is highly robust to small perturbations and variations in the system size, zero-mode topological boundary states can be probed via singular values even in finite-sized systems. Furthermore, the number of topological zero modes in the singular value spectrum is quantified by well-defined spectral winding numbers, thereby establishing a bulk-boundary correspondence that was limited to translationally invariant systems and the zero-mode edge state \cite{PhysRevLett.134.056601,f37j-l6wp}. On the other hand, since realistic systems are generally not perfect crystals, the study of amorphous and disordered systems
has also attracted widespread attention \cite{Zhang_2020,PhysRevResearch.2.012067,PhysRevLett.126.090402,PhysRevLett.127.213601,SciPostPhys.11.2.022,SciPostPhys.12.1.038,PhysRevB.106.014207,PhysRevLett.130.026202,qi2023localizationmobilityedgesnonhermitian,PhysRevB.107.024202,Zhang_2023,PhysRevLett.129.277601,PhysRevB.106.125310,Corbae_2023,SciPostPhys.15.5.193,SciPostPhys.17.3.086,He_2025,tp7r-31dj,hrbv-8jl1,steinfurth2026disorderimmunemomentumbandwinding,nl7n-bqn4,4z9l-63lr}.   Inspired by the combination of non-Hermitian topology and
amorphous systems, amorphous non-Hermitian topological phases have been proposed \cite{Manna_2023}. These studies can
give an abundance of materials hosting topologically protected states not restricted by translation symmetry and crystal symmetry. These phases hopefully promote further studies of potential applications of this field. However, due to the uncertainty of
the position of sites, the instability of energy spectrum
in such phases is more serious.  Therefore, the stable skin states are difficult to observe in the energy spectrum. How to reveal the point-gap topology and the corresponding skin states in these general systems is still lacking.

In this work, we address this challenge by developing a unified framework to characterize point-gap topology in one-dimensional amorphous non-Hermitian systems. Motivated by the intrinsic instability of the eigenvalue spectrum, we establish a correspondence between the number of stable zero-mode singular states of $H-E_b\mathbb{I}$ and the number of $E_b$-mode states of $H$ in the thermodynamic limit. Based on this correspondence, we construct a real-space topological invariant from the SVD of the enlarged Hamiltonian. This invariant accurately predicts the number of protected edge states and remains sharply quantized even in the presence of strong structural disorder. Our approach provides a general strategy for exploring point-gap topology in real space and redefines the non-Hermitian skin effect from a new perspective.

\begin{figure}[thb]
\centering
\includegraphics[width=1.02\columnwidth]{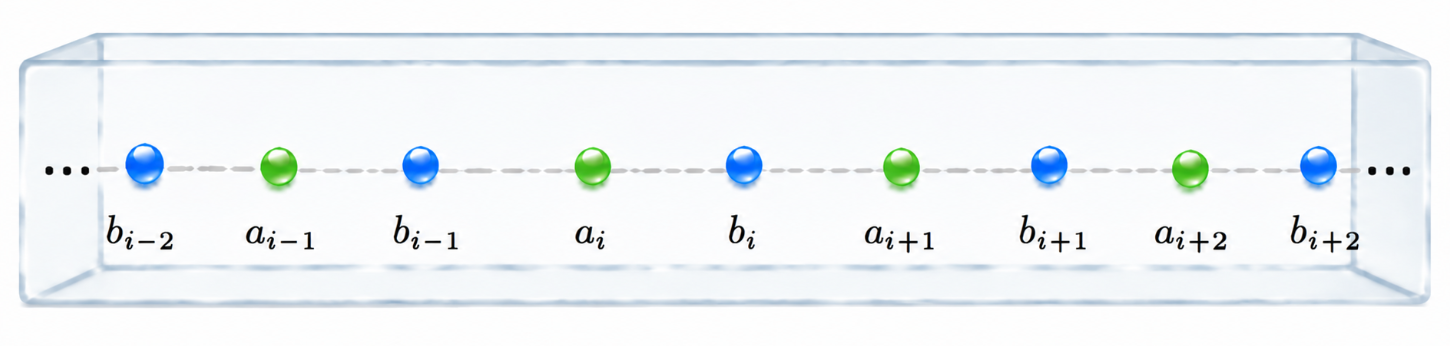}
 \caption{Schematics of the amorphous model on a chain.}
\label{fig1}
\end{figure}

\section{Model}
We consider a one-dimensional amorphous system with $2L$ sites randomly distributed in a region of length $L'$. In this work, we choose $L'=2L$ in all calculation. The site positions $\{x_{a_1},x_{b_1},x_{a_2},x_{b_2}......,x_{a_{L}},x_{b_{L}}\}$ are drawn from a random distribution over the interval $[0, L']$, with a minimum spacing constraint $d_{\min}$ to prevent site overlap. Specifically, we require that the distance between any two adjacent sites satisfies $\Delta> d_{\min}$. The Hamiltonian is given by
\begin{equation}
H=\sum_{i,j=1}^{L} t_{ij}(a^{\dag}_ib_j+\text{H.c.})+\sum_{k=1}^{L}\gamma(a^{\dag}_kb_k-b^{\dag}_ka_k),\label{Hmt}
\end{equation}
where $a_i^\dagger$ ($a_i$) and $b_j^\dagger$ ($b_j$) are creation (annihilation) operators for sublattices $a$ and $b$, respectively. We set $t_{ij}=te^{-S_{ij}/2}$, decaying exponentially to simulate realistic systems. The $S_{ij} = |x_{a_i} - x_{b_{j}}|$ is the distance between site $a_i$ and site $b_j$.  $t$ is the hopping amplitude and $\gamma$ is the nonreciprocal hopping. The schematic of the model is shown in Fig.~\ref{fig1}.

 \section{The invalidity of correspondence between non-Hermitian skin effect and spectral winding number } To characterize the localization properties and the non-Hermitian skin effect in the amorphous chain, we introduce several diagnostic quantities. The $n$-th normalized right eigenstate is $\psi_n=\sum_j \psi^{a}_n(j)\lvert j,a \rangle+\psi^{b}_n(j) \lvert j,b \rangle$, where $\lvert j,a\rangle$ and $\lvert j,b \rangle$ denote the state of the chain where the electron is on site $a_j$ and $b_j$, respectively. The inverse participation ratio (IPR) quantifies the degree of localization:
\begin{equation}
\text{IPR} = \left\langle \frac{1}{2L} \sum_{n=1}^{2L} \left[ \sum_{j} |\psi^a_n(j)|^4+|\psi^b_n(j)|^4 \right] \right\rangle,
\end{equation}
where $\langle \cdot \rangle$ denotes averaging over disorder realizations \cite{cheng2026unboundstatesmixedboundunbound}. An IPR of order $1/(2L)$ indicates fully extended states, while an IPR of order unity signals strong localization. 

However, a significant limitation of the IPR is that while it quantifies the degree of localization, it cannot differentiate between localization arising from Anderson mechanisms in the bulk and that induced by the NHSE at the boundaries. To address this issue, we employ the mean center of mass (mcom), which captures the spatial distribution of eigenstates. Specifically, the mcom is defined as
\begin{equation}
\text{mcom} = \left\langle \frac{\sum_{j=1}^{L} j \cdot \langle |\psi_n(j)|^2 \rangle_n}{\sum_{j=1}^{L} \langle |\psi_n(j)|^2 \rangle_n} \right\rangle,
\end{equation}
where $\langle |\psi_n(j)|^2 \rangle_n = \frac{1}{2L} \sum_n (|\psi^a_n(j)|^2+|\psi^b_n(j)|^2$) is the probability density averaged over all states \cite{lpm2-vcb4}. For a system with $2L$ sites, the geometric center is $L/2$; a significant deviation of the mcom from this value signals the NHSE.

While the mcom successfully identifies the spatial location of eigenstates, it does not directly quantify the degree of localization. To simultaneously capture both the localization strength and the boundary concentration of eigenstates in a single diagnostic quantity, we introduce the weighted inverse participation ratio (WIPR) \cite{f37j-l6wp}:
\begin{equation}
\text{WIPR} = \left\langle \frac{1}{2L}\sum_{n,j}( |\psi^a_n(j)|^4+|\psi^b_n(j)|^4)\cdot |j - \frac{L}{2}| \right\rangle,
\end{equation}
 The WIPR weights the IPR by the distance from the system center, thereby enhancing the contribution of states localized near the boundaries. A large WIPR indicates that the eigenstates are both localized and concentrated near the edges, which is the defining signature of the NHSE.

\begin{figure*}[tbh]
\centering
\includegraphics[width=2.05\columnwidth]{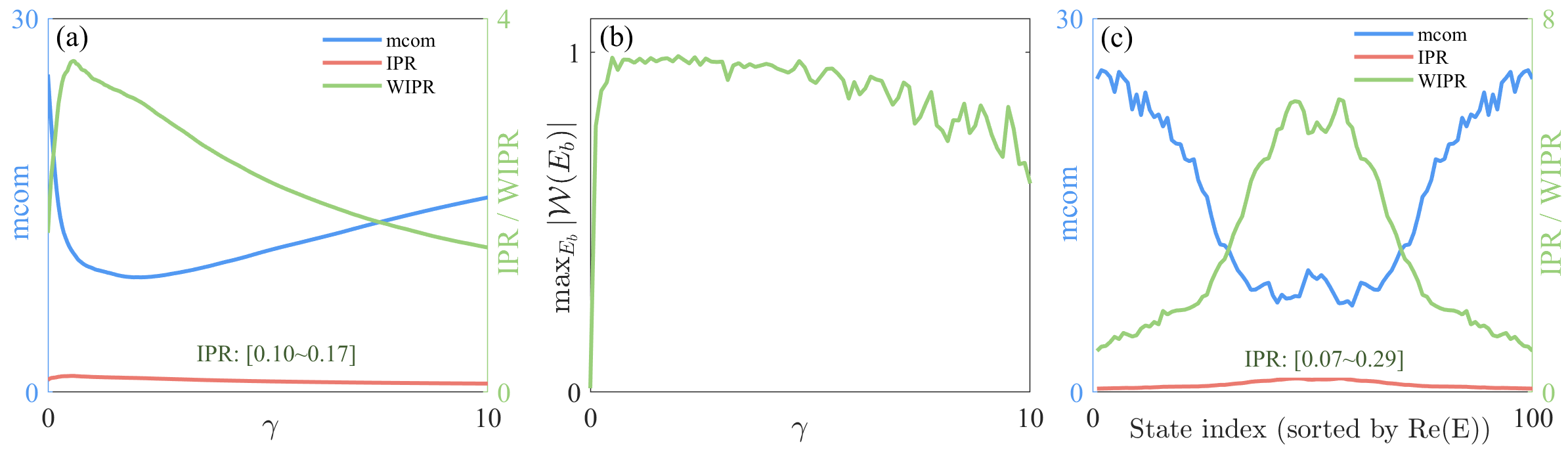}
 \caption{Characterization of the non-Hermitian skin effect via IPR, WIPR, mcom, and the real-space winding number for the amorphous chain ($t=1$). (a) Disorder-averaged mcom, IPR, and WIPR as functions of $\gamma\in[0,10]$ for $2L=100$ sites. (b) Maximum absolute value of the real-space spectral winding number $\max_{E_b}|\mathcal{W}(E_b)|$ as a function of $\gamma$ for $2L=100$ sites, averaged over $50$ disorder realizations per $\gamma$ point, with $E_b$ scanned over $\mathrm{Re}(E_b)\in[-5,5]$ and $\mathrm{Im}(E_b)\in[-(\gamma+1),\gamma+1]$. (c) Disorder-averaged mcom, IPR, and WIPR versus the state index (sorted by $\operatorname{Re}(E)$) at fixed $\gamma=0.2$ for $2L=100$ sites.}
\label{fig5}
\end{figure*}

Beyond these, the topological properties of the system can be characterized by a real-space winding number. When the determinant of $H-E_b\mathbb{I}$ is nonzero, there is a point gap at energy $E_b$. Following Refs.~\cite{PhysRevB.103.L140201,lpm2-vcb4}, the real-space spectral  winding number
can be expressed via the commutation relation
\begin{equation}
\mathcal{W}(E_{b})=\frac{1}{N^{\prime }}\text{Tr}^{\prime }\left( {Q}^{\dag }[{%
Q},{X}]\right) ,  \label{winding_comm}
\end{equation}%
where ${Q}$ is a matrix obtained from the
polar decomposition $H-E_b\mathbb{I}={Q}{P}$.
${Q}$ and ${P}$ are related to the SVD $H-E_b\mathbb{I}={U_A}{S}{U_B}^{\dag }$ via ${Q}={U_A}{U_B}%
^{\dag }$ and ${P}={U_B}{S}{U_B}^{\dag }$. ${X}$ is the
 operator with $X_{jj^{\prime },ss^{\prime }}=j\,\delta
_{j,j^{\prime }}\delta _{s,s^{\prime }}$ ($s$ can be sublattices $a, b$), and $\text{Tr}^{\prime }$
denotes the trace over the middle interval of length $N^{\prime }$,
excluding boundary contributions. If $\mathcal{W}(E_b)$ is non-zero, the system has NHSE.

The results of this analysis are summarized in Fig.~\ref{fig5}. Panel (a) shows that as $\gamma$ increases, the mcom drops sharply, signaling the onset of the NHSE, while the WIPR exhibits a pronounced peak at small $\gamma$, providing direct evidence of boundary-localized states. At larger $\gamma$, the WIPR gradually decreases and the mcom slowly recovers, indicating competition between the NHSE and disorder-induced Anderson localization. Panel (b) demonstrates that $\max_{E_b}|\mathcal{W}(E_b)|$ rises from zero to near unity at intermediate $\gamma$, confirming a topologically nontrivial point-gap phase, and gradually decays at strong non-Hermiticity. Panel (c) reveals that the WIPR peaks in the middle of the spectrum with complementary behavior to the mcom, indicating that the strongest NHSE manifests in the middle of the energy band.

However, in non-Hermitian disordered systems, the analysis of the traditional energy spectrum faces fundamental challenges. As pointed out in Ref.\cite{sirker2026pseudospectralphenomenaoriginnonhermitian}, the eigenvalue spectrum of a non-Hermitian Hamiltonian is highly sensitive to boundary conditions and general perturbations, making the spectrum itself an unstable object for encoding topological information. In particular, in amorphous systems lacking translational symmetry, this spectral instability is even more pronounced.

To quantitatively characterize this spectral instability, we introduce the concept of the pseudospectrum \cite{sirker2026pseudospectralphenomenaoriginnonhermitian,Trefethen2005}. For an operator $H$ and a positive parameter $\varepsilon$, the $\varepsilon$-pseudospectrum $\sigma_\varepsilon(H)$ is defined as the set of complex numbers $z$ satisfying
\begin{equation}
\lvert\lvert (z\mathbb{I}-H)^{-1} \lvert\lvert>\varepsilon^{-1},
\end{equation}
where $\|\cdot\|$ denotes the norm of operator and $\mathbb{I}$ is identity matrix. When $\|(z\mathbb{I} - H)^{-1}\| = \infty$, $z$ is an eigenvalue of $H$. Therefore, for any $\varepsilon > 0$, the spectrum is always contained within the $\varepsilon$-pseudospectrum. In this work,  we take the $\|\cdot\|$ to be 2-norm $\|\cdot\|_2$, $\sigma_\varepsilon(H)$ is equivalent to the set of complex numbers $z$ satisfying
\begin{equation}
s_{\text{min}}(z\mathbb{I}-H)<\varepsilon,
\end{equation} 
where $s_{\text{min}}(z\mathbb{I}-H)$ denotes the smallest singular value of $(z\mathbb{I}-H)$.

\begin{figure*}[tbh]
\centering
\includegraphics[width=2.05\columnwidth]{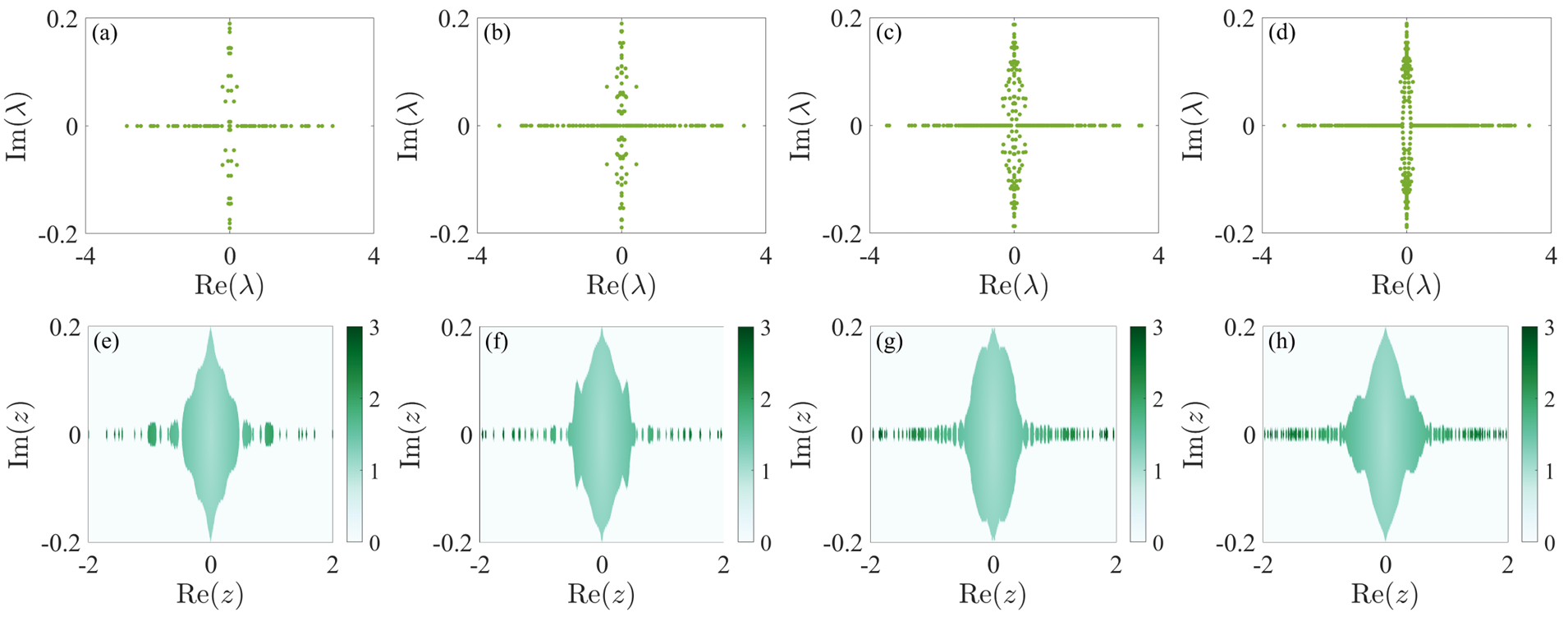}
 \caption{Eigenvalue spectra and pseudospectra of the $1$D non-Hermitian
amorphous chain defined by Eq.~\eqref{Hmt} for increasing system sizes. (a)--(d) Eigenvalues $\protect\lambda $ in the complex plane for systems
with $2L=100$, $200$, $300$, and $400$ sites, respectively. The hopping parameters are $t_{ij}=t\,e^{-|x_i-x_j|/2}$ with $t=1$,
$\protect\gamma=0.2$, and random positions $x_{a_i/b_i}\in \lbrack 0,2L\rbrack$ subject to a minimum spacing $d_{\min}=0.05$. (e)--(h) Corresponding $\protect%
\varepsilon $-pseudospectra with $\protect\varepsilon =10^{-2}$, where each dot marks $z\in \mathbb{C}$ satisfying $\protect s _{\min }(z%
\mathbb{I}-H)<\protect\varepsilon $, sampled on a grid $\mathrm{Re}(z), %
\mathrm{Im}(z)\in \lbrack -2,2\rbrack$ with spacing $0.005$. }
\label{fig2}
\end{figure*}

Fig.~\ref{fig2} illustrates the eigenvalue spectra and the corresponding $\varepsilon$-pseudospectra of the amorphous  chain for increasing system sizes. As the system size increases, the distribution of eigenvalues in the complex plane exhibits instability. The band structure becomes completely indistinguishable, in stark contrast to the well-defined discrete bands in crystalline systems. Comparing Figs.~\ref{fig2}(a)--(d) with Figs.~\ref{fig2}(e)--(h),  the pseudospectral regions extend far beyond the discrete eigenvalues and grow progressively larger as the system size increases from $L=50$ to $L=200$. More crucially, the pseudospectra reveal the extreme instability of the energy spectrum. This directly demonstrates that the topological bulk-boundary correspondence based on eigenvalues is practically unobservable in finite-size amorphous systems, consistent with the recent finding that point-gap topology fails to protect the skin effect in finite-size non-Hermitian systems~\cite{sirker2026pseudospectralphenomenaoriginnonhermitian}. This fundamental difficulty compels us to seek new approaches beyond traditional spectral analysis to reveal the topology of such systems.

\section{The robust point-gap topology in real space}
To reveal the topology of the system, we now introduce a connection between the zero-mode singular value and the emergence of mid-gap state of the energy spectrum in the thermodynamic limit. The SVD of the Hamiltonian reads $H-E_b\mathbb{I}=U_ASU^{\dagger}_B$, $E_b$ is a reference point on the complex plane \cite{PhysRevA.99.052118}. Denoting the column vectors of $U_B$ ($U_A$) by $ v_{n} $ ($ u_{n} $) and the singular values on the diagonal of $S$ by $s_{n}$, we obtain the equations:
\begin{equation}
(H-E_b\mathbb{I})^{\dagger }(H-E_b\mathbb{I}) v_{n} =s_{n}^{2} v_{n} ,\quad
\end{equation}%
and%
\begin{equation}
(H-E_b\mathbb{I}) v_{n}  =s_{n} u_{n} .
\end{equation}%
When $\lim_{ L\rightarrow \infty}s_{min}(H-E_b\mathbb{I})=0$, the $[(H-E_b\mathbb{I})  v_n]_{L\rightarrow \infty}=0 $ implies the system supports a state with wave function $v_n$ and eigenvalue $E_b$. Conversely, if $H$ has a stable  $E_b$-mode state with wave function $v_n$ in the thermodynamic limit, $[(H^{\dag}-E^{*}_b\mathbb{I}) (H-E_b\mathbb{I})v_n]_{L\rightarrow \infty}=0$ implies the emergence of a corresponding zero-mode state in the singular value spectrum of $H-E_b\mathbb{I}$. Therefore, the states of the energy spectrum in the thermodynamic limit can manifest in the singular value spectrum. Within this framework, the number of zero-mode singular values is directly linked to the number of stable mid-gap state in the energy spectrum. Crucially, the singular value $s_n$ of $H-E_b\mathbb{I}$ is the square root of the eigenvalue of the Hermitian operator $(H-E_b\mathbb{I})^{\dag}(H-E_b\mathbb{I})$. Consequently, singular values inherit the spectral stability of Hermitian matrices \cite{Trefethen2005,Ashida_2020}. Hence, the singular value spectrum is highly robust to small perturbations and variations in the system size, in stark contrast to the extreme instability of the eigenvalue spectrum demonstrated in Fig.~\ref{fig2}. This stability makes the singular value spectrum a powerful auxiliary tool for probing topological phases in non-Hermitian systems, even for finite-sized systems far from the thermodynamic limit~\cite{qy19-f9wm,f37j-l6wp}.

 However, whether the mid-gap states revealed by zero-mode singular values are genuine topological edge states must be further diagnosed by a suitable topological invariant. Here, we can construct the enlarged Hermitian Hamiltonian $\tilde{H}(E_b)$ to define topological invariant.  $\tilde{H}(E_b)$ is given by
\begin{equation}
\tilde{H}(E_b)={
\left[ \begin{array}{cccc}
0 & H-E_b\mathbb{I} & \\
H^{\dag}-E^{*}_b\mathbb{I} & 0 & \\
\end{array}
\right ]}.
\end{equation}
By construction, $\tilde{H}(E_b)$ possesses a chiral symmetry $\Sigma \tilde{H}(E_b) \Sigma^{-1} = -\tilde{H}(E_b)$ with $\Sigma = \text{diag}(\mathbb{I}, -\mathbb{I})$. This chiral symmetry endows $\tilde{H}(E_b)$ with a well-defined winding number that serves as the topological invariant for $H-E_b\mathbb{I}$:
\begin{eqnarray}
\mathcal{V}(E_b)=\frac{1}{2\pi i}\text{Tr} \ln (P^{A}{P^{B}}^{\dag}),\label{tvq3}
\end{eqnarray}
where the $P^{A}$ and $P^{B}$ are defined as $P^{S}=U_{S}^{\dag}\mathcal{P}U_{S}$ for $S=A,B$. Here, $U_{A}$ and $U_{B}$ are obtained from the SVD $H-E_b\mathbb{I}=U_{A} \mathcal{S} U^{\dag}_{B}$ with $\mathcal{S}=\text{diag}(s_1, s_2, \ldots, s_{2L})$ being the diagonal matrix of singular values~\cite{PhysRevB.103.224208,PhysRevLett.128.127601}. The matrix $\mathcal{P}$ encodes the real-space position information:
\begin{equation}
\mathcal{P}=\text{diag}[e^{-i2\pi \frac{1}{2L}}, e^{-i2\pi \frac{2}{2L}}, e^{-i2\pi \frac{3}{2L}},......].\\
\end{equation}
The topological invariant $\lvert \mathcal{V}(E_b) \lvert$ has a clear physical interpretation in that it counts the total number of topologically protected zero-mode edge states of the enlarged  Hamiltonian $\tilde{H}(E_b)$. Each edge state takes the form $\tilde{\psi}_n=\frac{1}{\sqrt{2}}(\tilde{\psi}^{c}_n,\tilde{\psi}^{d}_n)^T$, where $(H-E_b\mathbb{I})(H^{\dag}-E^*_b\mathbb{I})\tilde{\psi}^{c}_n=0$ and $(H^{\dag}-E^*_b\mathbb{I})(H-E_b\mathbb{I})\tilde{\psi}^{d}_n=0$. In the thermodynamic limit, if $\tilde{H}(E_b)$ possesses $\lvert 2\mathcal{V}(E_b) \lvert$ zero-mode edge states $\tilde{\psi}_n$ ($n=1,2,\ldots,2\mathcal{V}(E_b)$), then $H-E_b\mathbb{I}$ possesses exactly $\lvert \mathcal{V}(E_b) \lvert$ zero-mode singular values. Correspondingly, the original Hamiltonian $H$ hosts $\lvert \mathcal{V}(E_b) \lvert$ topologically protected edge states $\tilde{\psi}^{d}_n$ with eigenvalue $E_b$. This establishes a bulk-boundary correspondence in which the real-space winding number $\mathcal{V}(E_b)$, computed solely from the bulk SVD data, predicts the number of stable topological edge states at energy $E_b$ in the thermodynamic limit ~\cite{PhysRevLett.134.056601,qy19-f9wm,f37j-l6wp}. Compared to $\mathcal{W}(E_b)$, the remarkable advantage of this SVD-based topological invariant $\mathcal{V}(E_b)$ is its higher accuracy and less fluctuation in an amorphous system \cite{PhysRevB.103.224208}.

\begin{figure}[tbh]
\centering
\includegraphics[width=1\columnwidth]{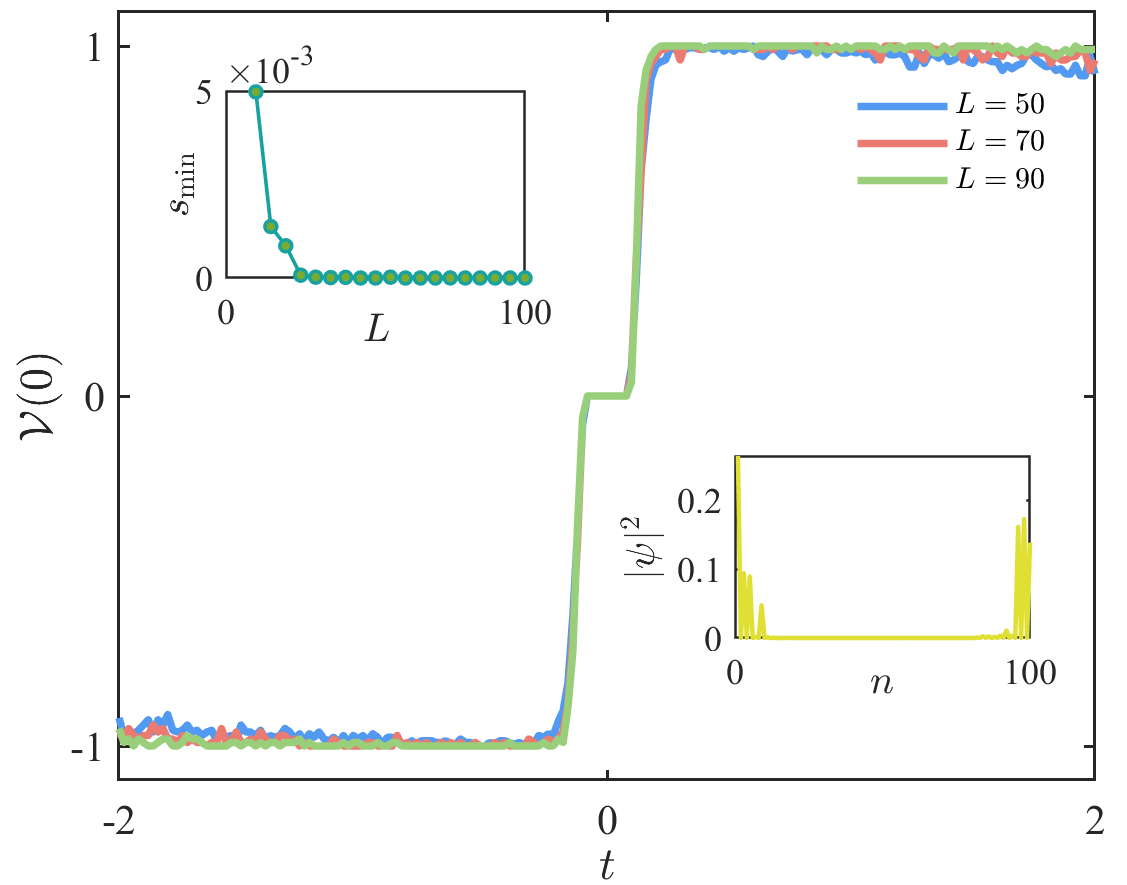}
 \caption{Winding number $\mathcal{V}(0)$ of the amorphous  chain as a function of the hopping amplitude $t$ for different numbers of sites $2L=100, 140, 180$, computed via the SVD-based invariant [Eq.~\eqref{tvq3}] with $\protect\gamma=0.2$, averaged over 1000 disorder realizations. Lower-right inset: spatial distribution of $\lvert \protect\psi\lvert^2=\lvert v_n \lvert^2$ for the smallest singular value at $t=-1.5$. Upper-left inset: the smallest singular value as a function of $L$ at $t=-1.5$.}
\label{fig3}
\end{figure}

\begin{figure}[t]
\centering
\includegraphics[width=0.92\columnwidth]{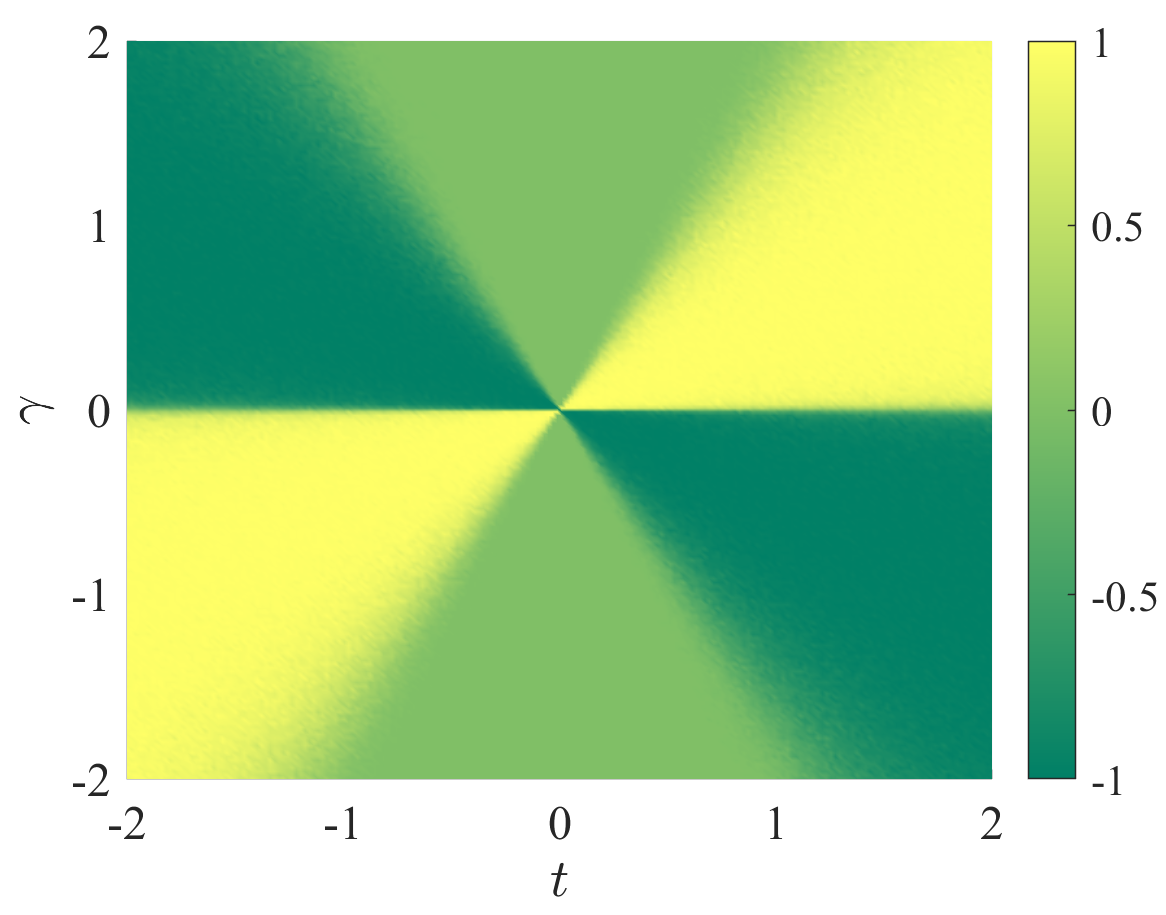}
 \caption{Topological phase diagram of the 1D non-Hermitian amorphous 
chain in the $(t,\protect\gamma )$ plane, obtained from the real-space winding number $\mathcal{V}(0)$. The system has $2L=100$ sites with random
positions $x_{a_i/b_i}\in \lbrack 0,100]$ and minimum spacing $d_{\min }=0.05$. The
result is averaged over $100$ independent disorder realizations for all $(t,\protect%
\gamma )$ points on a $200\times 200$ grid spanning $t,\protect\gamma \in
\lbrack -2,2]$. The color encodes the winding number $\mathcal{V}(0)$.}
\label{fig4}
\end{figure}

To observe the bulk-boundary correspondence in finite-sized systems with winding number $\mathcal{V}(E_b)$, we can calculate the $\mathcal{V}(E_b)$ zero-mode edge states in the singular values of $H-E_b\mathbb{I}$. We first focus on the case of $E_b=0$. Fig.~\ref{fig3} confirms that the winding number is sharply quantized, clearly distinguishing the topological ($\mathcal{V}(0)=\pm1$) and trivial ($\mathcal{V}(0)=0$) phases. The two insets verify that the zero-mode singular vector is localized at the chain ends, confirming the topological edge-state. Crucially, this edge state is entirely invisible in the eigenvalue spectrum due to severe spectral instability, yet it is unambiguously captured by the SVD-based approach. Fig.~\ref{fig4} presents the complete topological phase diagram of the amorphous  chain in the $(t,\gamma)$ parameter plane. The phase boundary between the topological ($\mathcal{V}(0)=\pm1$) and trivial ($\mathcal{V}(0)=0$) regimes is sharply resolved with a modest disorder averaging of 100 realizations per grid point, demonstrating the efficiency and robustness of the SVD-based invariant.  

\begin{figure}[h]
\centering
\includegraphics[width=1\columnwidth]{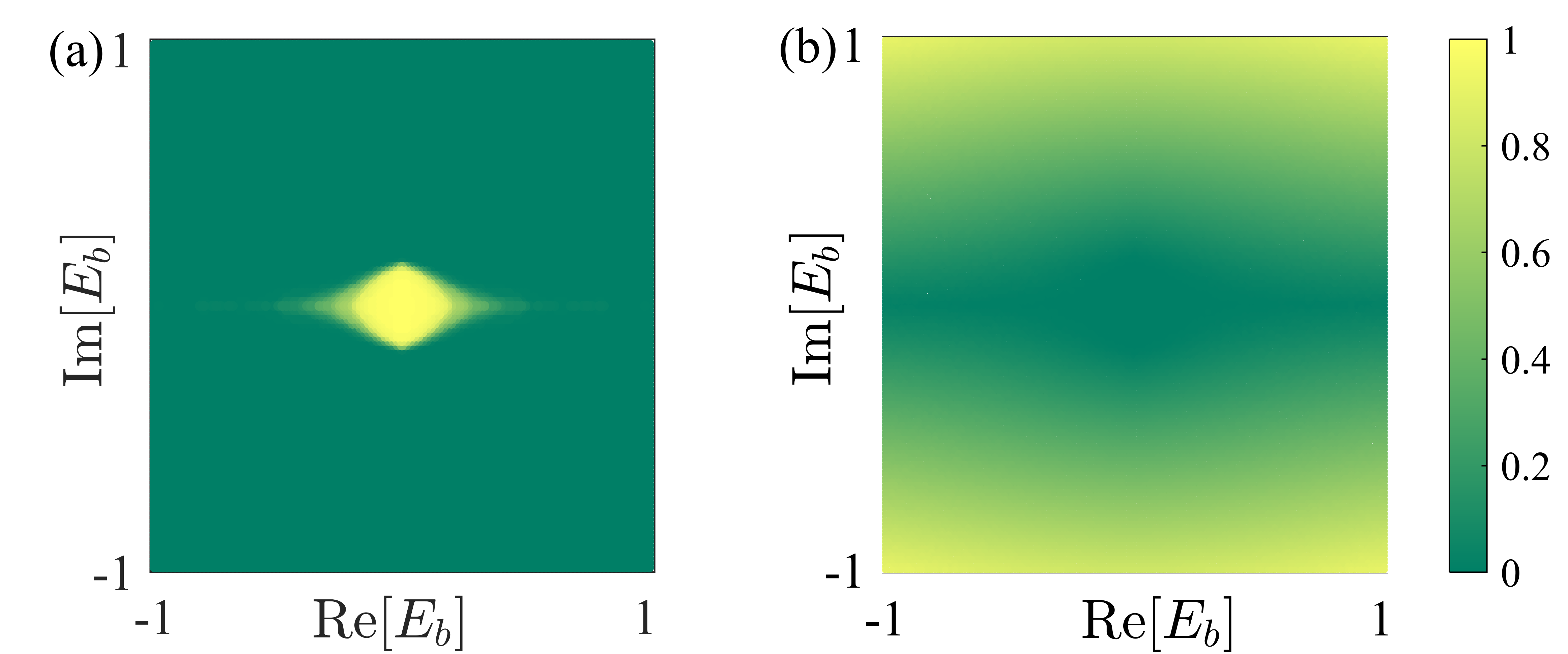}
\caption{(a) Real-space topological invariant $\mathcal{V}(E_b)$ and (b) the smallest singular value $s_{\min}$ of $H-E_b\mathbb{I}$ in the complex $E_b$ plane for the amorphous  chain with $t=1$, $\gamma=0.2$, $2L=140$, and $d_{\min}=0.05$, averaged over 100 disorder realizations.}
\label{fig7}
\end{figure}

To further obtain all the stable edge states of the system, Fig.~\ref{fig7} maps the winding number $\mathcal{V}(E_b)$ and the smallest singular value of $H-E_b\mathbb{I}$ across the complex plane of $E_b$. Panel (a) shows that $\mathcal{V}(E_b)$ is nonzero within a bounded region centered around $E_b=0$, delineating the topological point-gap regime. Panel (b) confirms that the region of zero-mode singular values coincides with the topologically nontrivial region identified in panel (a), demonstrating that zero-mode singular values faithfully signal the existence of topological edge states. As for the Fig.~\ref{fig7}(b), the zero-mode edge states in the singular value of $H-E_b\mathbb{I}$ are called skin states. The stable non-Hermitian skin effect can be observed in the singular values. Therefore, we redefine the non-Hermitian skin effect from a new perspective.

\section{Possible experimental realization}
The theoretical framework developed in this work can be verified on several established experimental platforms. The amorphous system with randomly positioned sites and exponential hopping may be engineered in waveguide arrays \cite{bk7q-6r9d}, photonic quasicrystal platforms \cite{PhysRevLett.132.263801}, or cold atoms where atoms are arranged via optical tweezers \cite{zhang2025observationnonhermitiantopologycold}. The topological edge states, invisible in the unstable eigenvalue spectrum, can be detected via the generalized Loschmidt echo $\mathcal{L}(\tau)=\langle \Psi | \Psi(\tau) \rangle / \sqrt{\langle \Psi(\tau)|\Psi(\tau)\rangle}$, where the initial state $|\Psi\rangle$ is a zero-mode singular state of $H-E_b\mathbb{I}$ and $\lvert \Psi(\tau) \rangle$ is the  evolutionary state at time $\tau$. For a topological edge state $\lvert \Psi \rangle$, $\mathcal{L}(\tau)$ remains close to unity, whereas for bulk states it decays rapidly, providing a clear dynamical fingerprint \cite{Naghiloo_2019,PhysRevLett.134.056601}. Based on these developments,
we believe that our approach is experimentally feasible.

 \section{Conclusion}In summary, we have developed a unified SVD-based framework to characterize the point-gap topology in one-dimensional amorphous non-Hermitian systems, establishing a correspondence between stable zero-mode singular values and mid-gap states in the thermodynamic limit. Since the singular spectrum is highly robust to small perturbations and variations in the system size, the bulk-boundary correspondence can be probed via singular values even in finite-sized systems. Our approach provides a general strategy for exploring point-gap topology in real space and redefines the non-Hermitian skin effect from a new perspective.

 \section{Acknowledgments}This work is supported by National Key Research and Development Program of China (No.2025YFE0103200), National Natural Science Foundation (Grants No. 12405007 and No. 12305011), Funds for Young Scientists of Chongqing Municipal Education Commission(Grant No. KJQN202400603 and No. KJQN202500619), Natural Science Foundation of Chongqing (Grant No. CSTB2025NSCQ-GPX1265, No. CSTB2022NSCQ-MSX0316, No. CSTB2024NSCQ-MSX0736, No. CSTB2025NSCQ-GPX1315, and No. CSTB2025NSCQ-GPX1272), and Chongqing Natural Science Foundation Project (Grant No. CSTB2025NSCQ-LZX0142)

\bibliography{references}
\end{document}